\documentclass[preprint,12pt]{elsarticle}

\usepackage{amssymb}
\usepackage{amsmath}

\usepackage{array}
\usepackage{booktabs}
\usepackage{caption}
\usepackage{float}
\usepackage{graphicx}
\usepackage{makecell}
\usepackage{multirow}
\usepackage{pdflscape}
\usepackage{siunitx}
\usepackage{tablefootnote}
\usepackage{tabularx}
\usepackage{lineno}

\begin{document}
\begin{frontmatter}



\title{Particle Inertia-Driven Pore Formation over Material Property Effects in Laser Powder-blown Directed Energy Deposition}

%

\author[a]{Dong Hee Kang} 
\author[b]{Samantha Webster} 
\author[a]{Sampson Canacoo} 
\author[c]{Samuel J. Clark} 
\author[c]{Kamel Fezzaa} 
\author[a]{Jihoon Jeong\corref{cor1}} 

\affiliation[a]{organization={Wm Michael Barnes '64 Department of Industrial \& Systems Engineering, Texas A\&M University},
            addressline={3131 TAMU}, 
            city={College Station},
            state={TX 77843},
            country={United States}}
            
\affiliation[b]{organization={Department of Mechanical Engineering, Colorado School of Mines},
            addressline={1610 Illinois St.}, 
            city={Golden},
            state={CO 80401},
            country={United States}}

\affiliation[c]{organization={X-ray Science Division, Argonne National Laboratory},
            addressline={9700 S. Cass Ave.}, 
            city={Lemont},
            state={IL 60439},
            country={United States}}            
            
\cortext[cor1]{Corresponding author: jihoonjeong@tamu.edu}

\begin{abstract}

Laser Powder-blown Directed Energy Deposition (LP-DED) offers flexibility for process and materials and high productivity ($\sim$5 kg/h), but process-induced pores often compromise mechanical properties. This study utilizes in-situ X-ray synchrotron imaging to compare pore formation mechanisms in Ti-6Al-4V (Ti64) and stainless steel 316L (SS316L), focusing on the interplay between particle dynamics and thermophysical properties. Four distinct pore formation mechanisms were identified, with most large pores originating from the closure of cavities formed behind incident particles impinging on the melt pool. High Weber number (We $\gg$ 1) governs this behavior, indicating that particle inertia, rather than thermophysical property differences, is the primary driver of large pore formation. The study demonstrates that increased energy density leads to larger melt pool volumes, facilitating deeper particle penetration. This greater penetration depth directly correlates with increased pore diameters. While thermophysical properties secondarily influence pore-formation frequency and cavity symmetry, particle inertia remains the dominant factor. These findings provide a physically grounded basis for understanding and controlling porosity in powder-blown DED processes.

\end{abstract}



\end{frontmatter}



\section{Introduction}
\label{sec1}

Laser powder-blown directed energy deposition (LP-DED) is one of the additive manufacturing (AM) techniques that has been a subject of attention in recent years due to its capability to manufacture parts with complex geometries, and repair parts used in aerospace and power industries \cite{ariasgonzalez21},\cite{gu21}. It also provides the ability to create parts with functionally graded materials, which would otherwise be impossible with traditional manufacturing methods \cite{piscopo22},\cite{saboori19}. LP-DED involves the use of a focused energy source to melt material deposited through a nozzle to build or repair parts by adding material layer by layer \cite{ahn21},\cite{gibson15}. 

Despite its advantages, parts fabricated using LP-DED usually contain pores generated during the process. These defects limit the application of LP-DED-fabricated parts by negatively affecting mechanical properties such as fatigue performance \cite{dang24}. Pores contribute to local stress concentrations, leading to the failure of parts \cite{svetlizky21}. It is therefore necessary to improve the understanding of the various pore formation mechanisms as a step in minimizing the occurrence of pores and producing higher quality parts.

X-ray imaging has been used frequently as a reliable technique to observe various phenomena in the AM process, including pore formation mechanisms and dynamics. The majority of these studies have been about keyhole instability in the laser powder bed fusion (LPBF) process \cite{zhao20}, \cite{cunningham19}. 
Martin et al. \cite{martin19_1} investigated pore formation mechanisms during laser powder bed fusion of Ti-6Al-4V alloy using in-situ X-ray imaging and a multi-physics simulation. They uncovered that pores form due to the collapse of the keyhole that is formed due to the deceleration and acceleration of the laser during turning. As the laser moves away from the point of turning and accelerates, the keyhole depression collapses due to instability. The inert shielding gas is trapped as molten metal fills the void, forming pores as the material solidifies. 
Huang et al. \cite{huang22_1}, using in-situ X-ray imaging during LPBF of the Al7A77 alloy, observed that pores form and grow rapidly when pinched off from the keyhole. These pores shrink as they move to the rear of the melt pool due to metal vapor condensation. Additionally, the pores may grow or shrink when captured by the solidification front. 
Sinclair et al. \cite{sinclair20_1} studied pore formation during single-track and multilayer printing of Ti-6Al-4V in overhang conditions. They identified multiple pore formation mechanisms, including trapped gas in the feedstock powder, hydrogen gas released from the decomposition of water vapor on the powder surface, and keyhole instability. Pores were also observed to form due to a lack of fusion, caused by insufficient laser penetration depth resulting from low energy density. 
Hojjatzadeh et al. \cite{hojjatzadeh19_1} studied the dynamics of pores during LPBF of AlSi10Mg. They saw that the pore movement behavior is governed by the competition of the thermocapillary force induced by the thermal gradient and the drag force from the melt flow. Under appropriate laser processing conditions, thermocapillary forces can effectively eliminate pores present in the feedstock powder.

Several in-situ X-ray imaging studies have been carried out on the LP-DED process to describe the cycle of pore formation, evolution, entrainment, and mitigation.
Wolff et al. \cite{wolff21_1} described how porosity formation in LP-DED differs from LPBF due to the influence of particle impact on keyhole dynamics and particle impact on melt pool flow.
Chen et al. \cite{chen21} observed that pores are formed due to argon gas present within the feedstock powder or hydrogen gas absorbed from the decomposition of moisture in the process chamber environment. 
Wang et al. \cite{wang22_1} discovered that pores are formed when spherical powder particles touch and rebound from the front keyhole region. Pores are also formed from the interaction of powder particles and melt flow. When a high-velocity particle is delivered to the melt pool, a cavity is formed in the path of the particle as it melts upon entry into the melt pool, leading to the formation of a pore when the shielding gas fills the cavity, and the open cavity closes. 
Bennett et al. \cite{bennett22_1} investigated the mitigation of powder-borne porosity. They observed a number of pore escape mechanisms, including particle melting, pore buoyancy, Marangoni flow, and the fluctuation of the liquid-vapor interface. They also observed some pore capture mechanisms to pore pinning by adjacent particles.
Webster et al. \cite{webster23} discovered a unique pore formation mechanism known as air cushioning, where gas is trapped between the incident particle and the melt pool surface.
K. Zhang et al. \cite{kzhang24_1} investigated that pore behavior can be divided into pore formation, bubble coalescence and growth, solid-liquid interface pushing of large bubbles, large bubble entrainment in the melt pool, and bubble escape or entrapment. Pores were attributed to gas present in the feedstock powder and existing pores from the substrate or previous track. They also stated that even though keyholes were observed in a few LP-DED studies, they do not generally apply to the LP-DED process since LP-DED is usually in conduction mode with a large laser spot size, with a large melt pool and a lower energy density than LPBF.
S. Zhang et al. \cite{szhang24_1} observed that pores are formed through six mechanisms. The unique pore formation mechanism is the split of a large pore into smaller-sized pores. Pores are also formed when a ripple-like surface wave occurs with the injection of powder particles into the melt pool. The other pore formation mechanism was found in the nucleation and growth of a pore after switching off the laser. This is speculated to be due to the hydrogen gas solubility difference between molten and solid metal. These studies have helped contribute to the understanding of pore formation mechanics. 
However, literature comparing pore formation in LP-DED in terms of the interplay between particle momentum and thermophysical properties remains limited. Although thermophysical properties can affect melt pool behavior, the relative contribution of particle inertia to pore formation has not yet been clearly established.

In this study, two materials used in LP-DED with remarkable differences in thermophysical properties, particularly density and surface tension gradient, were selected. The pore formation mechanisms triggered by incident particles on the melt pool were classified into four categories. The relative frequency of each mechanism and the pore size distribution were analyzed under identical energy density conditions. Utilizing an in-situ X-ray imaging technique, pore formation mechanisms of individual particles were captured for the instantaneous interaction between the particle inertia and the melt pool properties during the deposition process. To further our understanding, we observe interactions with previously deposited clads on multi-layer deposition, as this replicates AM scenarios where interactions between successive layers can influence pore formation and evolution. This approach enables a direct comparison of the relative roles of particle inertia and thermophysical properties in governing pore formation during LP-DED.

\section{Experimental}
\label{sec2}
\subsection{In-situ X-ray synchrotron imaging}
\label{subsec1}
Figure 1a shows the LP-DED setup used for in-situ synchrotron X-ray imaging, in which the X-ray beam penetrated the substrate to observe the interior of the melt pool during deposition. 
The LP-DED system was monitored in-situ X-ray synchrotron imaging setup at the 32-ID beamline, Advanced Photon Source (APS), at Argonne National Laboratory. The X-ray beam energy for imaging was set at 24.5 keV. The powders used were Ti-6Al-4V (Ti64, AP\&C) and 316L stainless steel (SS316, Oerlikon Metco), respectively. Both powders have a diameter distribution of 45 to 106 $\mu$m. 
The continuous wave laser with a 1070 nm wavelength (nLight AFX-1000) was used in the LP-DED system. The laser power is up to 1 kW, and the laser is focused on the surface of the substrate or the previously deposited layer with a Gaussian intensity profile beam mode. The laser power and scanning speed were regulated in a varied target range of energy density for $E=p/(v_s\cdot d)$, which is the generally adopted range in the LP-DED process for both materials (Supplementary Table S1). The laser power $p$ was set at 180, 210, and 240 W with $d$=150 $\mu$m beam diameter, and the scanning speed $v_s$ was varied at 10, 15, 20, and 25 mm/s. The process parameters for the calculated energy density conditions are described in Table 1.
In the commercial powder hopper (PF 2/2, GTV), the gas flow rate and disc speed were set to 10 LPM (liter per minute) and 3 RPM (revolution per minute), where the mass flow rate of powder was 34 g/min. The feedstock powder was delivered to the argon carrier gas through four single nozzles. The substrates were prepared with polished blocks measuring 30 mm in length, 10 mm in height, and 0.4 mm in thickness, and from materials of the same grade as the feedstock powders. Pore formation was described in the first and second layers (L1 and L2) for the deposition of both materials. X-ray synchrotron imaging captures the melt pool flow, incident particle diameter and velocity, and the moment of particle collision with the melt pool response.

\begin{figure}[h!]
\centering
\includegraphics[width=\textwidth]{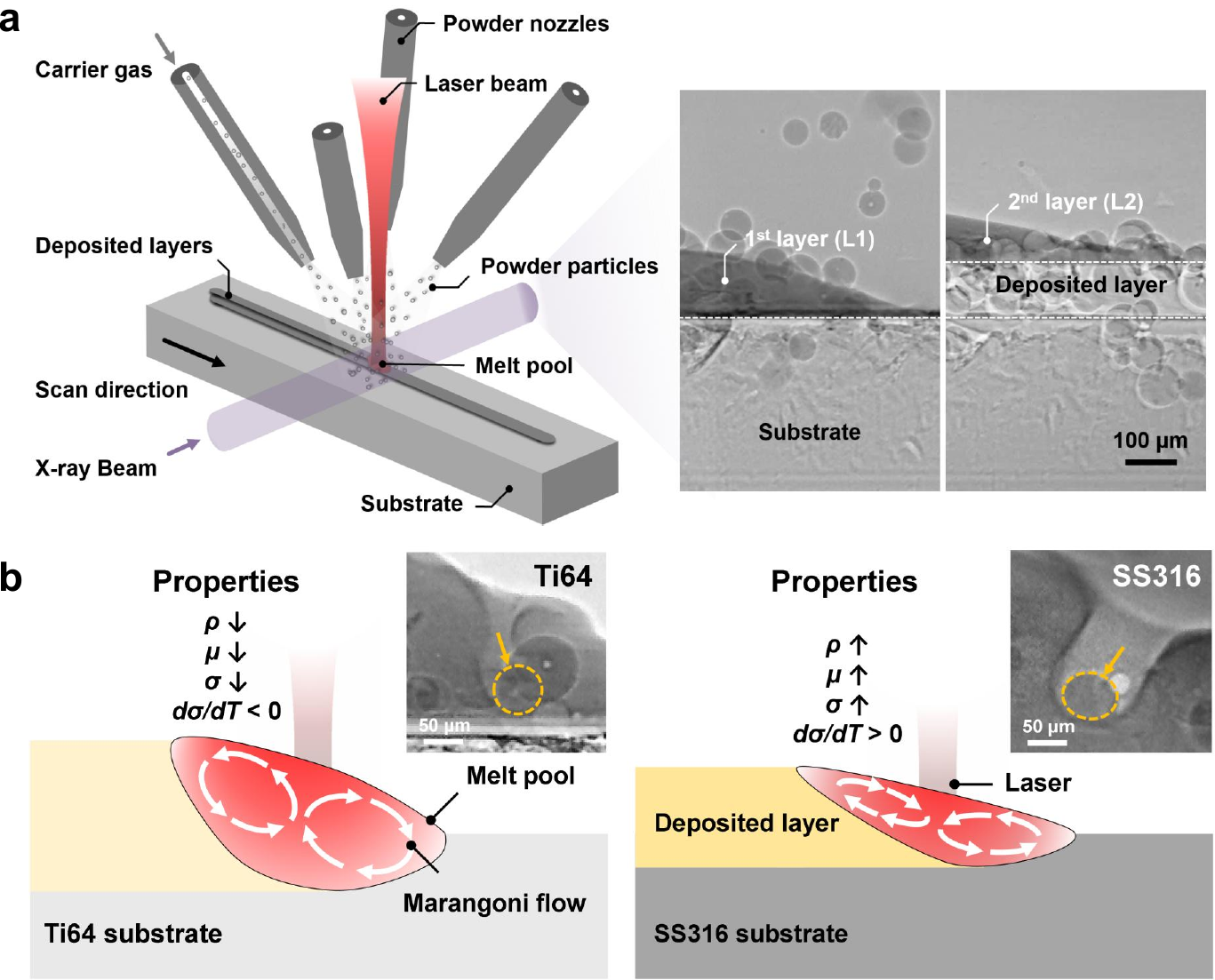}
\caption{Schematics of (a) the in-situ X-ray synchrotron setup for multi-layer deposition in LP-DED, and (b) visual comparison of the differences in melt pool geometry, relative thermophysical properties, and melt pool flow behavior between Ti64 and SS316 substrates.}
\label{fig1}
\end{figure}

\begin{table}[h!]
\centering
\caption{Process parameters for the Ti64 and SS316 powder deposition.}
\begin{tabular}{ccS}
\toprule
{\makecell{Laser power \\ (W)}} & {\makecell{Scan speed \\ (mm/s)}} & {\makecell{Energy density \\ (J/mm$^2$)}} \\
\midrule
\multirow{4}{*}{180} &  10&  120  \\
                     &  15&  80 \\
                     &  20&  60  \\
                     &  25&  48 \\ \cline{1-2}
\multirow{4}{*}{210} &  10&  140 \\
                     &  15&  93 \\
                     &  20&  70 \\
                     &  25&  56 \\ \cline{1-2}
\multirow{4}{*}{240} &  10&  160 \\
                     &  15&  107 \\
                     &  20&  80 \\
                     &  25&  64 \\
\bottomrule
\end{tabular}
\end{table}

\subsection{Image processing}
\label{subsec2}
The X-ray radiographs were obtained for 700 frames per layer deposition and repeated twice for each sample energy density condition. Particle collision was observed at a 50 kHz (20 $\mu$s per frame) frame rate of the radiographs (Photron, Fastcam SA-Z) with 2.0 $\mu$m per pixel size from a 100 $\mu$m thick LuAG:Ce scintillator. Using the Fiji open-source platform, sequential images were divided into an average image, $J_{avg}$, which was taken 100 frames before layer deposition of each sample \cite{schindelin12}. The raw X-ray radiographs were denoted as $I_i$, where $i$ ranges from 1 to 700, which were divided by $J_{avg}$ to obtain images according to the equation, $I_i'=I_i/J_{avg}$, to get a clear melt pool boundary. 

\section{Results and Discussion}
\label{sec3}
Pore formation occurs both directly and indirectly through the interaction between particle inertia and melt pool dynamics. Depending on its relative location, a pore within the disturbed melt pool may either escape through the melt pool surface or become trapped at the solidification front. Within the melt pool, whose size is on the order of the beam diameter, a steep temperature gradient develops, giving rise to Marangoni flow. In addition, irregular particle impacts vigorously disturb the melt pool flow. The following sections describe the underlying formation mechanisms and the resulting pore behavior.

The high-speed camera operating at 50 kHz captured melt pool flow and particle impact dynamics during multilayer deposition. 
The image resolution was sufficient to resolve the melt pool flow as well as the interaction of individual particles with the melt pool, as shown in Fig. 1b, providing a level of observation comparable to previous studies (Supplementary Table S2). We categorized pore formation mechanisms into four states for a melt pool interacting with particles and describe sequential images of the interaction in Figs. 2 to 5. 
To clarify the pore formation moment, colored dashed lines were added. The dashed lines depict boundaries of the melt pool interface (green), substrate (white), powder particle (orange), pore (red), and solidification front (blue). Arrows along the boundaries indicate changes in these dashed line contours and depict position variations.

\subsection{Pore formation mechanisms}
\label{subsec1}
\begin{figure}[h!]
\centering
\includegraphics[scale=0.6]{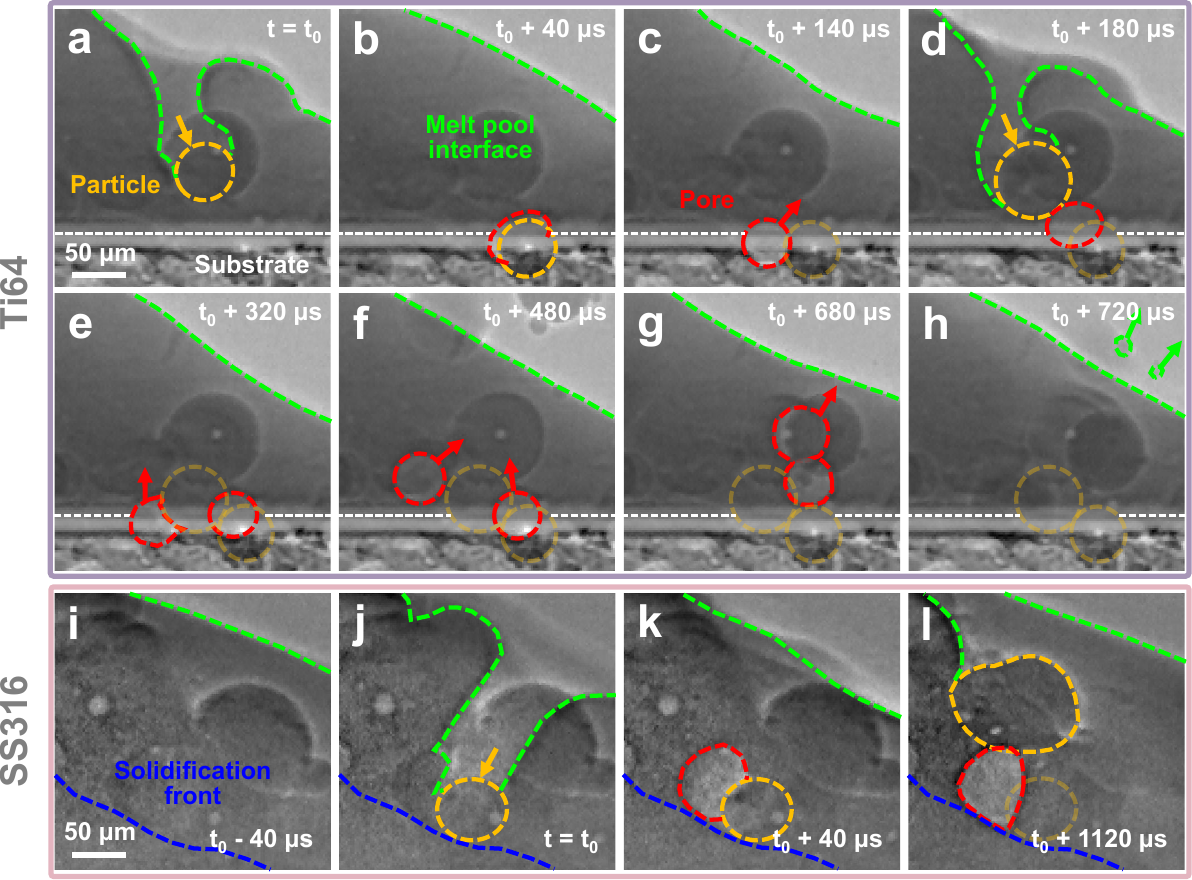}
\caption{First pore formation mechanism, M1: Gas pore entrainment from the cavity closure of the melt pool interface by a particle collision. Large cavity formation behind an injected particle of (a-h) Ti64 and (i-l) SS316 to the melt pool.}
\label{fig2}
\end{figure}

The first mechanism of pore formation, M1, describes gas pore entrainment from the cavity closure of the melt pool interface by a particle collision in Fig. 2. In Fig. 2a-h, images of the pore formation by fluctuation of the Ti64 melt pool interface. An injected particle into the melt pool deforms the interface, leaving a large cavity trailing the penetration direction compared with the particle size. The sequential images were sufficient to capture the cavity closure procedure. The pore minimizes its surface area within the melt pool on the impinged particle surface at the solidification front. The sequentially injected particle collides with the trapped particle. The second pore is entrained within a melt pool, then both pores rise. At the head of the melt pool, a steep temperature gradient extends from the melt pool interface to the solidification front. Marangoni stresses arising from temperature gradients generate recirculation in the melt pool, carrying pores along with the flow. Bursting out pores at the interface could form spatters in Fig. 2h. The sequential pore formation process by M1 occurred within 1 ms, which was noticed by the large pore formation with the high-speed camera. Supplementary Video S1 depicts particle penetration into the Ti64 melt pool and the pore formation of M1. Under the inert gas flow, in contrast, a melt pool interface without particle supply from the feedstock does not create cavities and pores, as shown in Supplementary Video S2. The role of incident particles is evident when the two videos are compared. The large pore formation is consistent with M1 driven by particle impact on the melt pool interface.
Within the observed time window, most pores escape the melt pool out of the interface. However, the pore could be confined by other injected particles into the melt pool. A pore is confined at the solidification front due to another particle collision over the pore in Fig. 2i-l. The penetrated particle impinges on the solidification front of the SS316 melt pool in Fig. 2l. In the solidification front at the tail of the melt pool, the particle's surrounding thermal energy was dissipated by heat conduction. When the particle did not acquire sufficient heat to melt completely, the pore became trapped with partially melted state. In Supplementary Video S3, particle penetration into the SS316 melt pool depicts the various pore formation mechanisms. Some of the generated pores become trapped at the tail of the solidification front.

Some incident particles reach the edge of the melt pool, located away from its center. In other words, a cavity is not sufficiently developed behind the incident particle, since the melt pool is not deep enough for the particle to be submerged. Nevertheless, some particles may sustain a thin gas layer on their surface as they slip into the melt pool. 
The second pore formation mechanism, M2, is the most frequently observed pore formation mode and occurs either when an incident particle interacts with a shallow melt pool or when gas is trapped along the particle surface within the melt pool, as shown in Fig. 3. 
The pore contraction occurred on an injected particle surface within the melt pool. Irregular particle surfaces induce more frequent and larger pore volumes \cite{wolff21_1}. A thin gas layer on the surface of a particle exists within the carrier gas due to the viscosity of the gas. At the solidification front, the thin gas layer surrounding a trapped particle contracts into discrete gas pores, thereby minimizing gas volume. As the particle surface near the pores begins to melt, pores around the molten particle could float. The pores flow together along buoyancy and the Marangoni flow.

\begin{figure}[h!]
\centering
\includegraphics[scale=0.6]{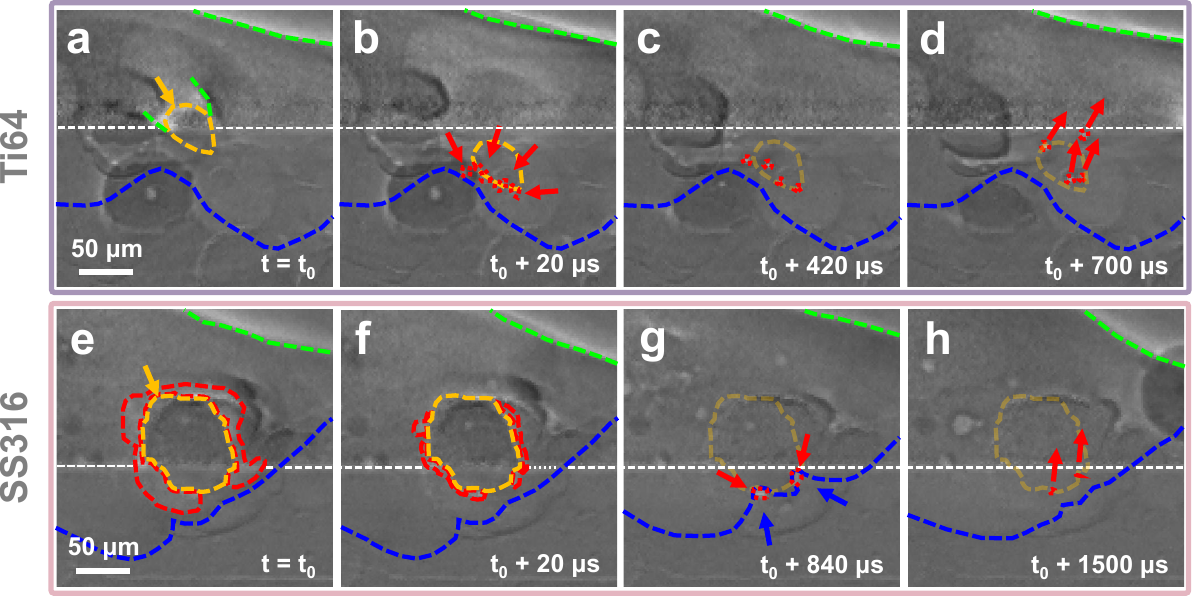}
\caption{Second pore formation mechanism, M2: Gas pore contraction from the boundary of particle surface within a melt pool. Contraction of a thin inert gas layer on the particle surfaces of (a-d) Ti64 and (e-h) SS316.}
\label{fig3}
\end{figure}

\begin{figure}[h!]
\centering
\includegraphics[scale=0.6]{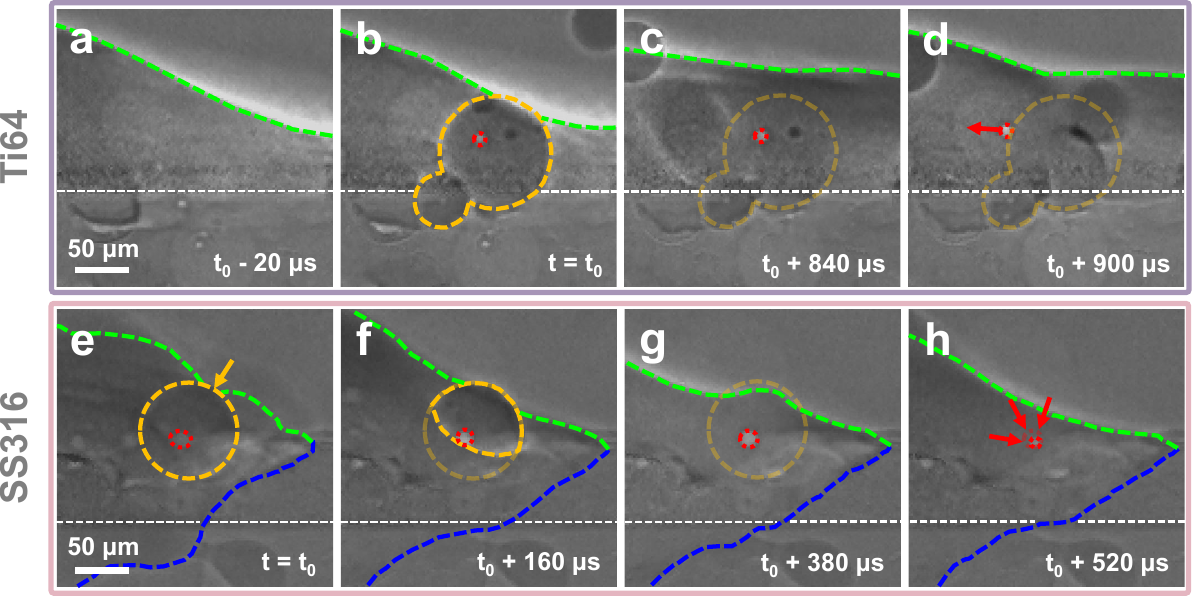}
\caption{Third pore formation mechanism, M3: Gas pore release from the inherent pore in the particle. Internal pores in the (a-d) Ti64 and (e-h) SS316 particles are released when the material surrounding the pores is completely melted.}
\label{fig4}
\end{figure}

The third pore formation mechanism, M3, describes the pores originating from inherent particle powder. Regardless of collisions, the release of a pore entrained within a particle occurs only once complete melting of the particle has taken place, as shown in Fig. 4.
Upon exposure to the melt pool, the pore contracts to a pressure equilibrium state. Then the pore could float after the particle fully melted \cite{bennett22_1}. In Fig. 4e-h, the boundary of the melt pool interface is changed as the particle melts as time goes on. Pore release requires more time than other mechanisms, depending on the particle size and the location of the pore in the particle. 

\begin{figure}[h!]
\centering
\includegraphics[scale=0.6]{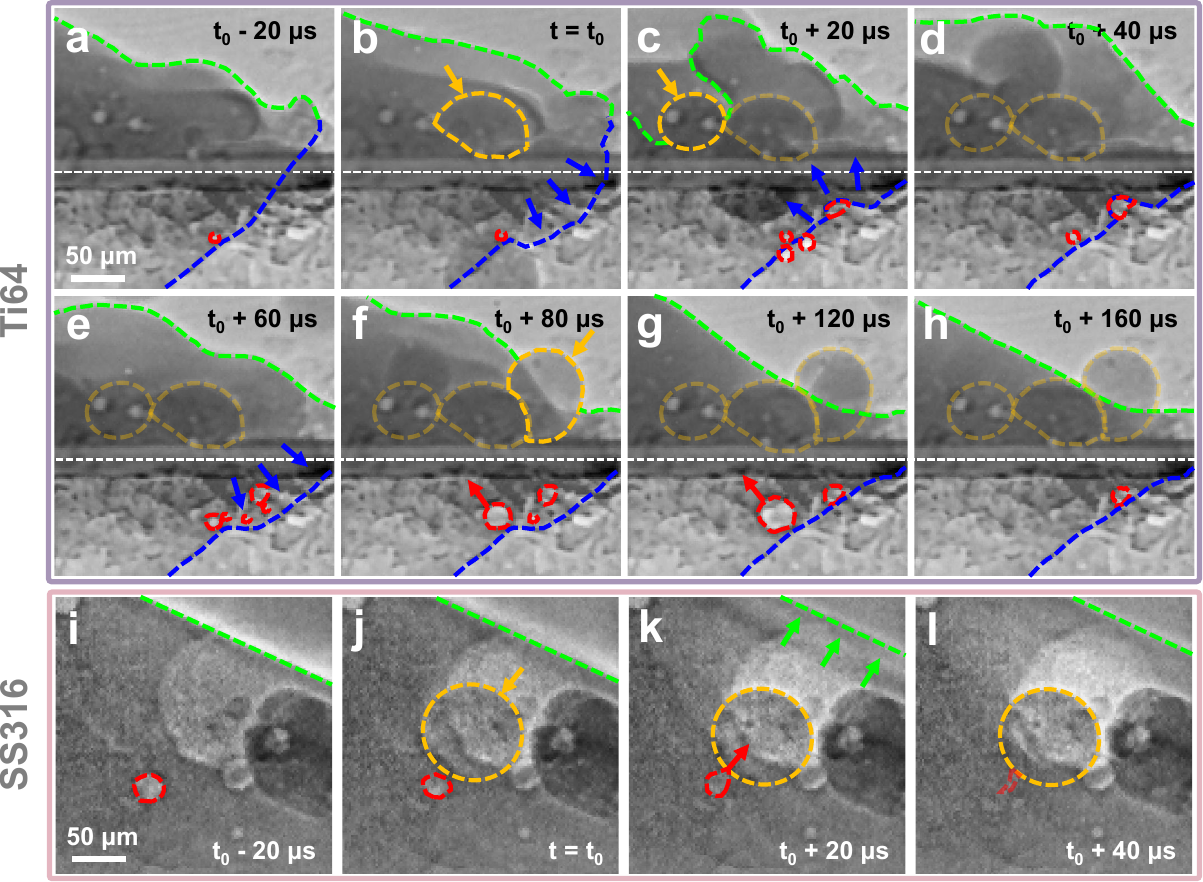}
\caption{Fourth pore formation mechanism, M4: Gas pore discharge from the solidification front by a particle impact into the melt pool. Discharge of the trapped gas pore at the solidification front in the previous layer of (a-h) Ti64, and (i-l) SS316  due to stimulation by particle collision into the melt pool body.}
\label{fig5}
\end{figure}

The mechanisms of pore formation describe the direct interaction of the particle collision with the melt pool interface (M1 and M2) and the inherent pore release from the particle (M3). Meanwhile, the fourth pore formation mechanism M4 describes that the particle collision stimulates the submerged pores at the solidification front in Fig. 5. 
Pores within the solidification front seem to have originated from the lack of fusion of particles under the substrate/previous layer. In M4, the impact of a particle on the melt pool modifies the borderline of the solidification front suddenly. As shown in Fig. 5d-f, several pores expanded simultaneously and subsequently merged due to melt pool motion induced by a surface wave generated by the impact of another particle. The merged pore floats out to the melt pool or ruptures in the melt pool. If the pores could not escape to the melt pool, they contract back down below the solidification front, as in Fig. 5b-d. In Supplementary Video S4, waves appear to spread out across the interface of the melt pool from the point of particle impact. Also, in the SS316, particle collision stimulates the pore at a solidification front in Fig. 5j-l. The collision of a particle generates a pressure wave, which could lead to cavitation becoming pores due to an instantaneous pressure difference in a melt pool. Especially, the momentum of a particle could deform the boundary of the solidification front, creating a gas channel where a cluster of particles has a part of the lack of fusion. The flow of molten metal to the pore along the gas channel fills out the pore. The particle collision with the solidification front can contribute to pore elimination during layer deposition by the momentum of a particle. The trapped pore at the solidification front can be discharged for tens of $\mu$s. 
The other incident particles bounce off the solidification front of the melt pool. The particles collide with a particle, which is exposed to an interface in the shallow melt pool. Then the early submerged particle is imprisoned in the melt pool. Some particles bounce off the melt pool interface, which is not enough to generate cavities or rupture pores. 

In summary, pore formation in LP-DED was classified into four mechanisms: inertia-driven cavity closure gas entrainment (M1), gas entrapment along the particle surface (M2), gas release from inherent pores within the particle (M3), and reactivation of pores trapped beneath the solidification front by particle impact-induced pressure field changes (M4). All of these mechanisms were observed by in-situ X-ray imaging in both materials, demonstrating that the pore formation stage is governed primarily by particle inertia rather than by differences in thermophysical properties.

\subsection{Melt pool geometry}
\label{subsec2}
The melt pool geometry is depicted depending on the energy density. To examine the effects of material type and deposited layer number, the melt pool geometry was quantified in terms of melt pool length, depth, and angle between the melt pool interface and the solidification front, as illustrated in Fig. 6a.

\begin{figure}[h!]
\centering
\includegraphics[width=\textwidth]{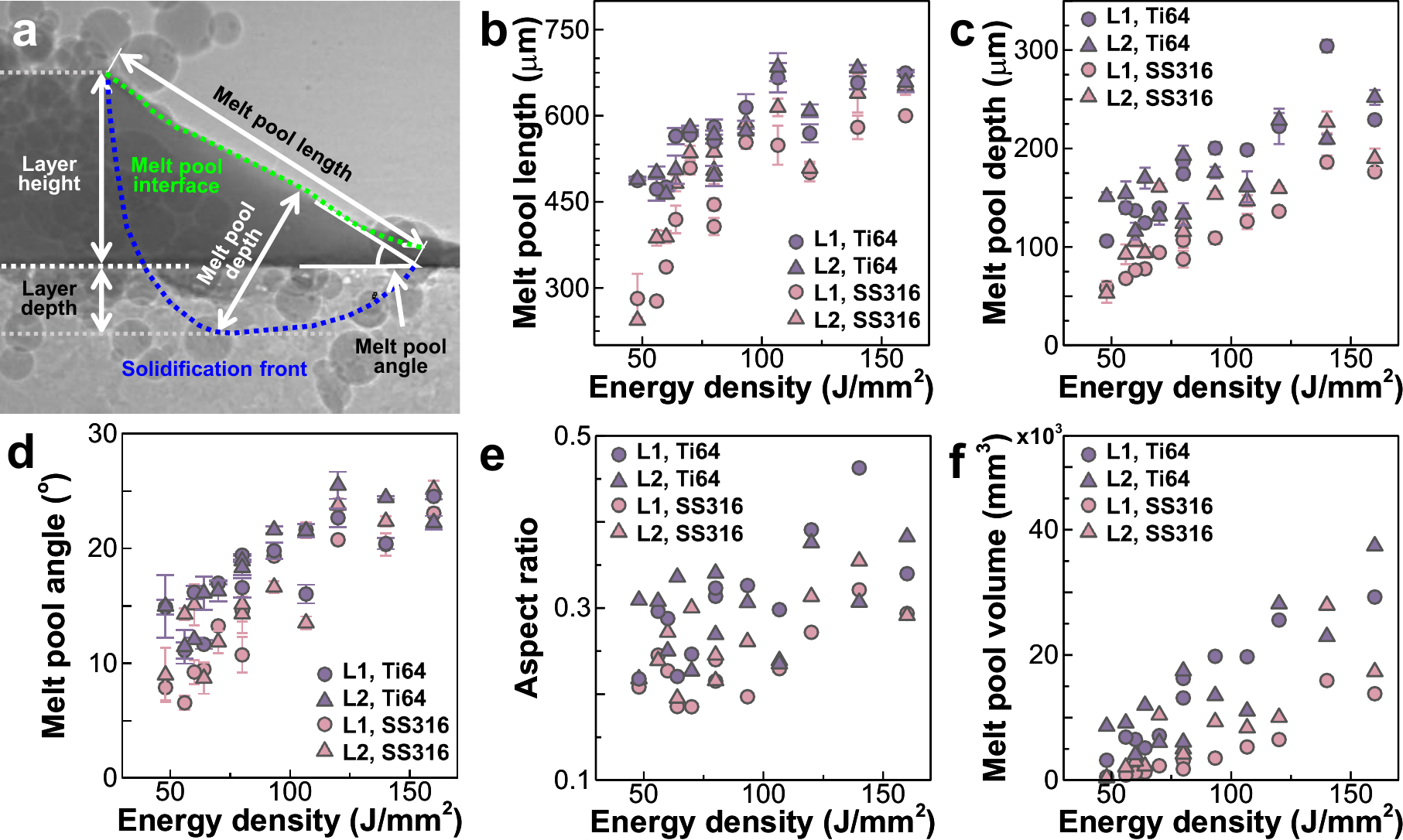}
\caption{Geometric dimensions of melt pool for Ti64 and SS316 depending on the energy density. (a) Schematic of a melt pool's dimensions during layering deposition. Analysis of geometric dimensions for (b) length, (c) depth, and (d) angle of melt pool. (e) Aspect ratio of the melt pool for its depth to length ratio. (f) Volume of melt pool by melt pool dimensions.}
\label{fig6}
\end{figure}

\begin{table}[h!]
\caption{Thermophysical properties of the Ti64 and SS316.}
\begin{tabularx}{\linewidth}{p{6.0cm}X X}
\toprule
Properties & Ti64 \cite{tseng19} & SS316 \cite{le19} \\
\midrule
\multirow{2}{*}{Density, $\rho$ (kg/m$^3$)} & 4420 (at r.t.$^*$) & 7950 (at r.t.) \\
& 3920 (at $T_l$) & 6881 (at $T_l$) \\
Liquidus temperature, $T_l$ (K) & 1923 & 1723 \\
Solidus temperature, $T_s$ (K) & 1877 & 1658 \\
Dynamic viscosity, $\mu$ (kg/m$\cdot$s) & 3.25$\times$10$^{-3}$ (at $T_l$) & 5.6$\times$10$^{-3}$ (at $T_l$) \\
Surface tension, $\sigma$ (N/m) & 1.525 (at $T_l$) & 1.710 (at $T_l$) \cite{fukuyama21} \\
Surface tension gradient, $d\sigma$\slash$dT$ (N/m$\cdot$K) & -0.28$\times$10$^{-3}$ & 0.52$\times$10$^{-3}$ \cite{fukuyama21} \\
\multirow{2}{*}{Thermal conductivity, $k$ (W/m$\cdot$K)}  & 5.9 (at r.t.) & 13.4 (at r.t.) \\ 
& 27.9 (at $T_l$) & 28.5 (at $T_l$) \\
\multirow{2}{*}{Specific heat, $c$ (J/kg$\cdot$K)}  & 547 (at r.t.) & 470 (at r.t.) \\ 
& 831 (at $T_l$) & 830 (at $T_l$) \\
\bottomrule
\end{tabularx}
\footnotesize{$^*$r.t.: room temperature, 298 K}
\end{table}

The melt pool length is the straight line distance between the edges of the melt pool interface. The melt pool depth is the longest perpendicular distance away from the tangent of the solidification front to the melt pool length. The melt pool angle is the angle between the line of the melt pool length and the line of the substrate/previous layer. The aspect ratio is calculated from the melt pool depth-to-length ratio. The melt pool volume \textit{V} is assumed to be a part of the elliptical cap. The \textit{V} is calculated as below\cite{chen21}, 
\begin{equation}
V=\frac{\pi W}{3(L/2)}D^2 \left (3\frac{L}{2}-D \right)\\    
\end{equation}
The \textit{D} is the melt pool depth, and the width of the melt pool \textit{W} is assumed identical to the melt pool length \textit{L}. 
For all dimensions, the melt pool geometries of Ti64 were larger than those of SS316. Because the liquid density of SS316 is approximately 80\% higher than that of Ti64, the melt pool volume of SS316 remained smaller, particularly in the low energy density regime. As shown in Fig. 6b, the melt pool length in SS316 increased rapidly at low energy density and then more gradually with further increases in energy density. The melt pool depth scaled approximately linearly with energy density, while the difference between the two materials remained nearly constant at about 100 $\mu$m over the full range, as shown in Fig. 6c. The melt pool angle also gradually increased with energy density, as shown in Fig. 6d. Under conditions of sufficient particle supply, this increase in melt pool angle suggests that incident particles were more likely to become submerged in the melt pool rather than rebound from the surface. This interpretation is consistent with the observed increases in both melt pool depth and melt pool length at higher energy densities.
The aspect ratio in Fig. 6e describes the normalized melt pool shapes, showing that the melt pool becomes shallower at lower energy densities and that SS316 is generally shallower than Ti64. At the same energy density, melt pool volume differs by up to three times between the materials in Fig. 6f. These contrasts indicate that differences in material properties strongly influence melt pool development despite the same energy input. 
Under identical energy densities, SS316 has a higher liquid density and nearly comparable specific heat. SS316 has a substantially larger volumetric heat capacity than Ti64, requiring more energy (75\% higher than Ti64) to heat and sustain an equivalent molten volume. In addition, its higher thermal conductivity at room temperature promotes stronger heat dissipation into the substrate and previously deposited layers, further suppressing melt pool growth despite its lower liquidus temperature.

\subsection{Relative frequency of pore formation}
\label{subsec3}
High-speed in-situ synchrotron X-ray imaging enables the categorization of pore formation mechanisms occurring on the sub-millisecond timescale during particle interaction with the melt pool. In some cases, multiple mechanisms of pore formation can occur simultaneously, and each mechanism was counted separately. The relative frequency of pore formation by four mechanisms is analyzed in Fig. 7. The pore formation is classified depending on the mechanisms that contribute to pore formation. Relative frequency is defined as the probability of pore formation associated directly with particles interacting with the melt pool, including both contacting and entering particles. The relative frequency of pore formation is compared to the energy density, material (Ti64 and SS316), and deposition layer (L1 and L2).

\begin{figure}[h!]
\centering
\includegraphics[width=\textwidth]{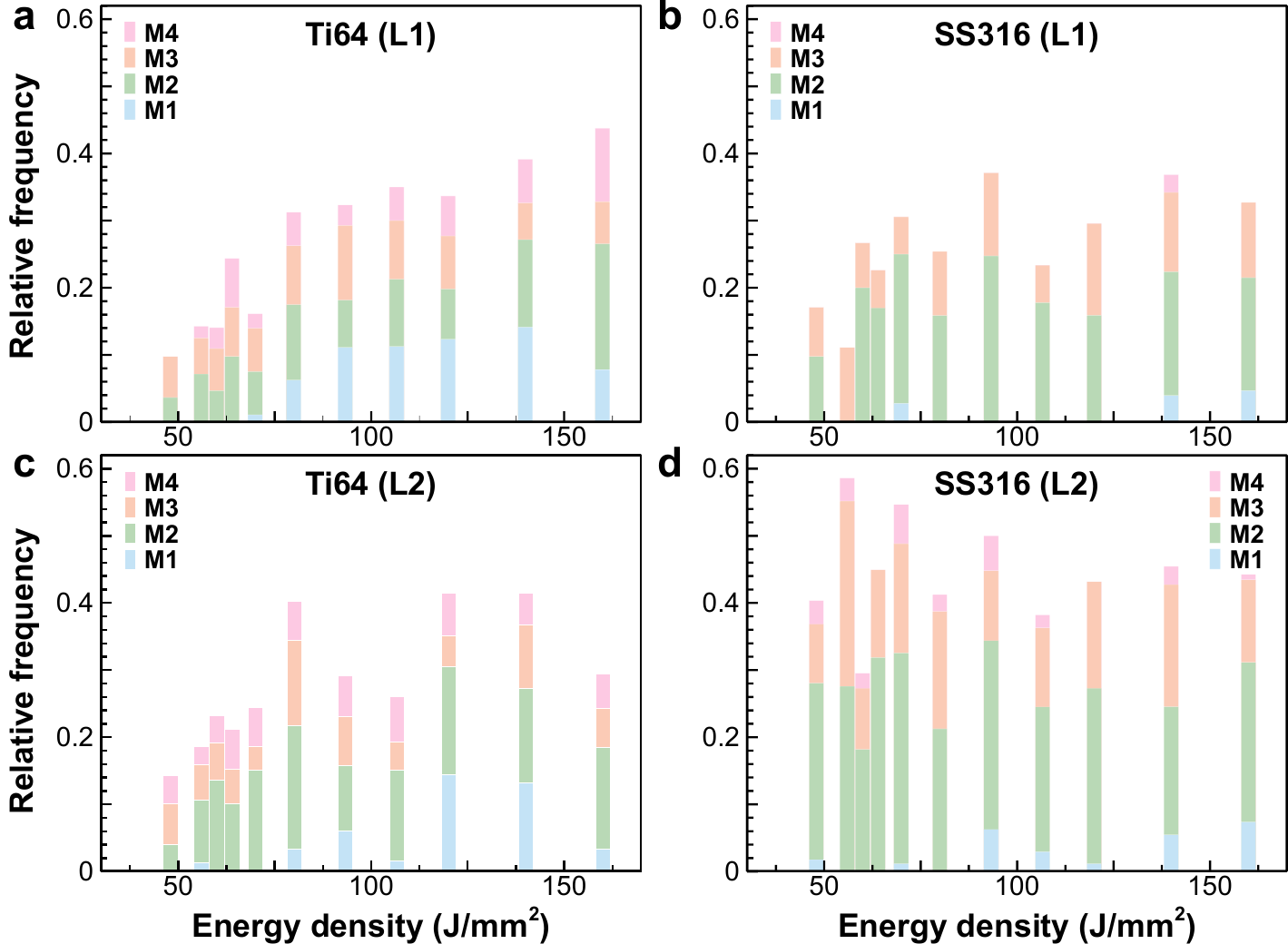}
\caption{Relative frequency of pore formation with four mechanisms by the injected particles to melt pool at first and second layer (L1 and L2) deposition of (a,c) Ti64 and (b,d) SS316 depending on the energy density.}
\label{fig7}
\end{figure}

In Ti64 depositions (L1 and L2), the relative frequency of pore formation increases as energy density rises, predominantly due to an increased incidence of M1. This behavior is consistent with the findings in melt pool geometry in Fig.6f, which showed that the increase in melt pool volume with energy density raises the probability of pore formation through cavity closure in M1. 
Conversely, M1 occurred only rarely in both layers of the SS316 melt pool. The relatively small depth and lower aspect ratio of the SS316 melt pool indicate that the melt pool was not sufficiently deep, relative to particle size, to develop a cavity behind the incident particle. Consequently, pore formation directly induced by particle impact was predominantly governed by M2 in SS316, as Fig. 7b.

In Fig. 7d, the increased relative frequency of pore formation in L2 of SS316 under low energy density conditions is attributed to insufficient remelting and incomplete melt pool development in this regime. Consistent with the melt pool length trend of SS316 in the low energy density regime shown in Fig. 6b, a larger fraction of incident particles is likely to remain partially melted under these conditions. As a result, particles interacting with the less fully remelted previously deposited surface are more likely to trap gas along the particle surface, thereby increasing the occurrence of M2. In addition, insufficient remelting at low energy density can promote the lack of fusion regions, which is consistent with the appearance of M4 in this regime. In SS316, the relative frequency of M4 was higher in L2 (2.6\%) than in L1 (0.2\%). In addition, M3 accounted for a substantial fraction of pore formation in both materials, with average frequencies of 7.2\% and 6.3\% for L1 and L2 in Ti64, and 9.1\% and 14.6\% for L1 and L2 in SS316, respectively.

\subsection{Pore formation characteristics}
\label{subsec4}
Pore formation characteristics are analyzed by the mean size of pores generated according to the four mechanisms. Fig. 8a shows the distribution of pore diameter by the pore formation mechanisms. Box plots display the pore diameter distributions, including quartiles for each mechanism. The plots are presented in a vertically split panel. Ti64 appears on the left and SS316 on the right for each mechanism. The lower whisker corresponds to the minimum value. The upper whisker is defined as $Q_3 + 1.5 IQR$. Values of pore diameter beyond the upper threshold are described as outliers in the pore formation distribution. The median values of pore diameter distribution for M1, M2, and M4 are higher for the SS316 than those of the Ti64.
Under nearly identical mass flow rate conditions, the variation in incident particle counts between the two materials agreed with that expected from their density difference. The resulting counts were 52$\pm$3 per 1 ms for Ti64 and 20$\pm$11 per 1 ms for SS316. The number of supplied particles with the same powder size distribution is inversely proportional to their densities. 

\begin{figure}[h!]
\centering
\includegraphics[width=\textwidth]{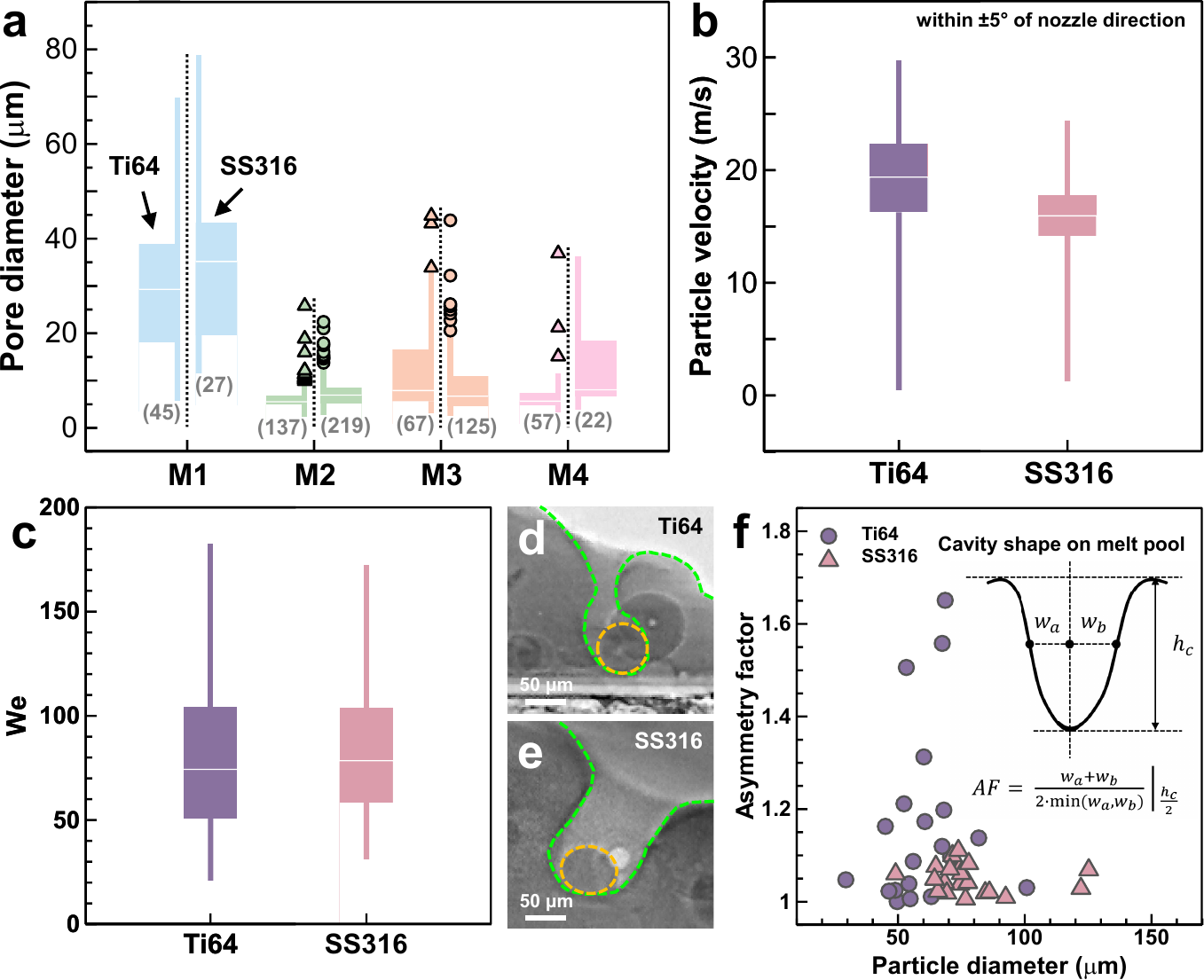}
\caption{(a) Pore diameter within melt pool by the four pore formation mechanisms for Ti64 and SS316 particles. The gray numbers below the bar charts indicate the number of cases analyzed. (b) Particle velocity within the $\pm$5$^\circ$of the nozzle direction. (c) Weber number of the Ti64 and SS316 particles generated by M1.
Cavity formation behind the particle penetration to (d) Ti64 and (e) SS316 melt pool. (f) The asymmetry factor (AF) describes the symmetrization of cavity shape with the particle diameter.}
\label{fig8}
\end{figure}

Although the number of pores generated by M1 is lower than that of other mechanisms, M1 is the major region generating large pores. The median of M1 for the Ti64 and the SS316 are 29.1 and 35.2 $\mu$m, respectively. The medians for the other mechanisms of pore formation are $<$ 10 $\mu$m, especially, the M2 is the most frequent pore formation mechanism with the narrow $IQR$ and the smallest minimum pore diameter. Outliers observed in M2 may stem from ambiguities in the criteria used to distinguish M2 from M1. In the M3, the pore diameter within a melt pool is independent of particle kinetic energy. For M3, the median pore diameter is less than 8 $\mu$m. When a particle containing a pore strikes the tail of the solidification front, the resulting pores can cause remaining defects in the deposition layer unless they are subsequently remelted. Because the inherent pore within the particles takes more time to float from the particle compared with other mechanisms, which require melting out the particle surrounding the pore.
The discharged pores associated with M4 were larger in the SS316 melt pool. While the onset of M4 is governed by particle impact-induced disturbance, the resulting pore size appears to be modulated by material thermophysical properties. The relatively higher density, dynamic viscosity, and surface tension of SS316 may contribute to maintaining larger and more stable pore diameters within the melt pool.
For M1, M2, and M4, the median pore diameter was consistently larger in the SS316 melt pool than in Ti64. While pore generation was dominated by particle inertia, the larger pore size observed in SS316 suggests a secondary influence of thermophysical properties. The higher dynamic viscosity of SS316 may act to retard pore deformation and collapse, thereby promoting the persistence of larger pore diameters within the melt pool.

The box plots in Fig. 8b illustrate the velocity distributions, with the box indicating the interquartile range of first to third (Q1-Q3), the horizontal line denoting the median, and the whiskers extending to the minimum and maximum values. The maximum velocity of the particles is up to 29.8 m/s and 24.2 m/s for Ti64 and SS316, respectively. For the same gas flow rate of the powder feeder, the mean velocity of Ti64 particles is 21.6\% higher than that of SS316 for the same LP-DED process.  

The depth by particle penetration to the melt pool, forming a volume of cavity, is assumed to be proportional to the magnitude of particle momentum. Weber number, $We = {\rho v^2l}/{\mu}$, was introduced to understand the physical characteristics of cavity formation at the melt pool interface. In this dimensionless form, $v$ and $l$ represent the particle velocity and the characteristic particle diameter, respectively. The \textit{We} number describes the ratio of inertial force to surface tension during particle impact onto the liquid interface and is commonly used to characterize splashing and cavity formation phenomena during liquid-solid interaction \cite{kintea16, techet11}.
To account for the statistical variability in particle diameter, particle velocity, and melt pool thermophysical properties, the \textit{We} number was evaluated using a Monte Carlo framework based on the propagation of input distributions. 
The sampled inputs included the experimentally measured particle diameter and velocity distributions, together with temperature-dependent liquid density and surface tension evaluated over the physically bounded melt pool temperature range from the liquidus to the vaporization temperature using the property relations summarized in Supplementary Tables S3 and S4. Detailed information on the assigned input distributions, sampling bounds, and the number of Monte Carlo trials is provided in the Supplementary Note S1. The resulting \textit{We} distributions for both materials are presented as box plots, where the box bounds represent Q1 and Q3 and the whiskers denote the 95\% coverage interval. The median values were 74.34 and 78.45 for Ti64 and SS316, respectively, indicating that both materials exhibit a comparable \textit{We} range, as shown in Fig. 8c.

This result suggests that, within the physically bounded LP-DED melt pool temperature range, the variation in \textit{We} is governed primarily by the dispersion in particle velocity. Under melt pool conditions in an Ar-purged chamber, particle inertia is more significant than the temperature-dependent changes in liquid density and surface tension in governing the formation of large pores.
Although density and surface tension vary over the melt pool temperature window, their influence on \textit{We} remains secondary because of the strong quadratic dependence on particle velocity. Since the \textit{We} number directly reflects the balance between inertial impact and surface tension resistance, the large \textit{We} values observed for both materials (\textit{We} $\gg$ 1) indicate that particle inertia dominates cavity formation. 
Therefore, the formation of large cavities in M1 is primarily attributed to high-velocity particle impact during LP-DED, with temperature-dependent thermophysical property changes playing a secondary role under the present conditions.
This interpretation is physically consistent with the experimentally observed splash-assisted cavity formation behavior in both Ti64 and SS316 melt pools, as shown in Fig. 2.

The asymmetry factor (AF) describes the cavity shape of a melt pool by an incident particle in Fig. 8d,e. The asymmetrically closed cavity could develop into small pores.
In other words, an asymmetric cavity trailing the incident particle may close earlier than a more symmetric one, thereby forming a smaller pore. The AF suggests that differences in cavity symmetrization are associated with material-dependent melt pool dynamics, with viscosity likely playing an important role.
For the Ti64 and SS316 melt pools, 20 cases of cavity formation caused by incident particles (M1) were analyzed. The AF is calculated as below, 
\begin{equation}
AF=\frac{w_a+w_b}{2\cdot \text{min}(w_a,w_b)}\Bigg|_{\frac{h_c}{2}}\\
\end{equation}
For an image containing a cavity, the cavity height ($h_c$) can be quantified. Specifically, $h_c$ is the distance from the particle edge at the melt pool interface to the interface centerline, evaluated at both cavity ends. Half widths ($w_a$,$w_b$) are perpendicular distances at the half maximum point of $h_c$ to the interface boundary. In all still images used for the AF analysis, $h_c$ was required to be greater than twice the particle diameter at the melt pool interface. Prior to touching the solidification front, the particle had to remain within the melt pool. Figure 8f shows the AF as a function of incident particle diameter. SS316 exhibited a lower AF distribution than Ti64. An AF value of 1 indicates a perfectly symmetric cavity shape at the melt pool interface.

The AF distributions reveal a clear material-dependent difference in cavity symmetry. Ti64 exhibited a broader distribution, with AF values extending up to 1.65, whereas SS316 showed a narrower distribution concentrated at lower AF values, with a maximum near 1.11. Because lower AF values correspond to a more symmetric cavity shape, the cavity trailing the incident particle in SS316 is inferred to remain larger and more stable before closure. Such delayed closure is consistent with the entrainment of larger pores. This interpretation agrees with the experimentally observed M1 pore size distribution, in which the pores formed in SS316 were larger than those in Ti64. The lower AF values in SS316 may be associated with material-dependent melt pool dynamics, with viscosity likely contributing to the preservation of a more symmetric cavity shape. The material-dependent variation in AF is therefore expected to influence the size distribution of large pores formed through M1.

\begin{figure}[h!]
\centering
\includegraphics[width=\textwidth]{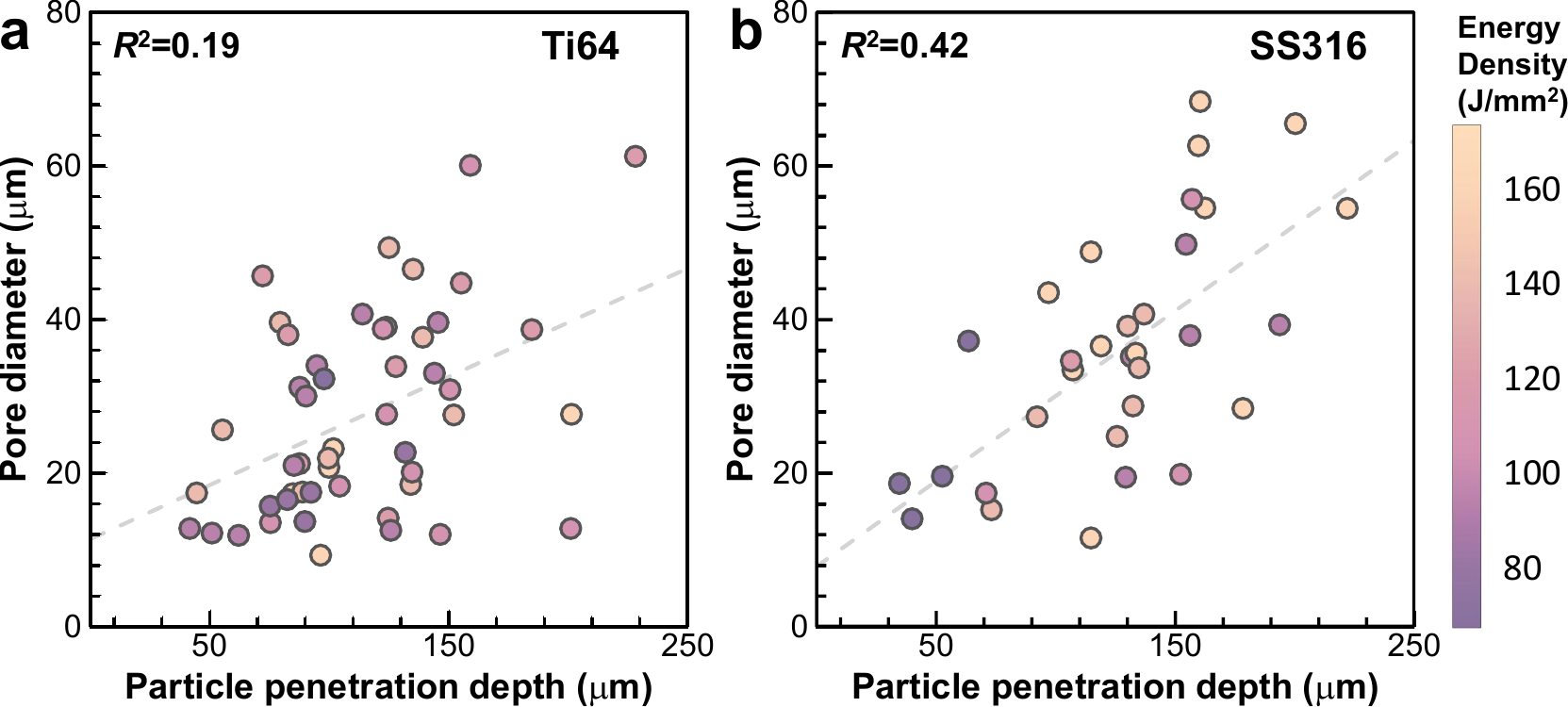}
\caption{Pore diameter as a function of particle penetration depth into the melt pool for M1 under different energy densities for (a) Ti64 and (b) SS316.}
\label{fig9}
\end{figure}

The pore diameter distribution as a function of particle penetration depth for M1 is shown in Fig. 9. The pore diameter was positively correlated with particle penetration depth. As the energy density increased, the melt pool became deeper, allowing greater particle penetration and, consequently, the formation of larger pores. The plotted data also show that higher energy densities corresponded to both larger accessible penetration depths and larger pore diameters.
However, the strength of this correlation differed between the two materials. In Ti64, the relationship between penetration depth and pore diameter exhibited relatively large scatter, resulting in a weak linear correlation ($R^2=0.19$). In contrast, SS316 showed a stronger correlation with lower variance about the fitted trend ($R^2=0.42$). This difference is consistent with the AF analysis in Fig. 8f. In Ti64, more frequent early cavity closure is expected to broaden the pore diameter distribution regardless of penetration depth, leading to greater scatter. By contrast, the more symmetric cavity shape in SS316, corresponding to AF values closer to 1, suggests that cavity closure occurs later and more consistently, so that pore diameter is more directly proportional to particle penetration depth.

As a result, the in-situ synchrotron X-ray observations demonstrate that pore formation in LP-DED is governed primarily by particle inertia during particle-melt pool interaction. The image resolution enabled direct identification of four pore formation mechanisms. It confirmed that high velocity incident particles ($v_{\textrm{mean,Ti64}}$= 19.3 m/s, and $v_{\textrm{mean,SS316}}$= 15.8 m/s) generate large cavities under conditions of \textit{We}$\gg$ 1. In this regime, the formation of large pores is primarily driven by inertia-induced cavity development and particle penetration into the melt pool. As the melt pool becomes deeper with increasing energy density, greater particle penetration becomes possible, which in turn promotes the formation of larger pores. Although both materials exhibited comparable \textit{We} ranges, their pore-size distributions differed because thermophysical properties secondarily influenced melt pool geometry and cavity evolution. In particular, the AF analysis showed that SS316 maintained a more symmetric cavity shape than Ti64, consistent with delayed cavity closure and the formation of larger M1 pores. By contrast, the broader AF distribution in Ti64 indicates more frequent early cavity closure, resulting in a wider scatter of pore diameter and a larger population of relatively small pores even at similar penetration depths.

\section{Conclusion}
\label{sec4}
Pore formation during LP-DED was directly observed by in-situ synchrotron X-ray imaging, enabling the interaction between incident particles and the melt pool to be resolved on the sub-millisecond timescale. The observations show that pore formation occurs both directly and indirectly through particle-melt pool interaction and can be classified into four mechanisms: direct cavity closure entrainment by particle impact (M1), gas entrapment along the particle surface in a shallow melt pool (M2), release of inherent pores from the particle itself (M3), and reactivation of pores trapped beneath the solidification front by particle impact-induced disturbance (M4).
The dominant factor governing large pore formation is particle inertia rather than thermophysical property differences. Under the present LP-DED conditions, both Ti64 and SS316 exhibited large Weber numbers (\textit{We} $\gg$ 1), indicating that the impact of incident particles generated cavity formation in an inertia-dominated regime. The depth of particle penetration into the melt pool was strongly correlated with pore diameter, and increasing energy density promoted deeper penetration by increasing melt pool depth and volume. Therefore, larger pores were formed at greater penetration depths.

Thermophysical properties nevertheless played an important secondary role by modifying melt pool geometry and, consequently, the relative frequency of pore formation mechanisms. The melt pool of SS316 remained smaller and shallower than that of Ti64 under identical energy density conditions. M1 occurred less frequently in SS316, whereas M2 became the dominant pore formation. In multilayer deposition, insufficient remelting and the roughness of the previous clad further increased the occurrence of M2 and M4, particularly under low energy density conditions.
Although both materials showed comparable \textit{We} ranges, differences in thermophysical properties influenced cavity evolution after impact. The AF analysis showed that SS316 tended to maintain a more symmetric cavity shape, consistent with delayed cavity closure and the formation of larger M1 pores. Whereas Ti64 showed broader AF distributions and more frequent early cavity closure, resulting in a wider scatter of pore diameter. Thermophysical properties mainly affected the geometry of the melt pool, the occurrence frequency of each mechanism, and the subsequent evolution of pores, while the primary origin of large pore formation remained 
particle inertia. 

These findings clarify that pore formation in LP-DED should be understood primarily from the viewpoint of inertia-driven particle impact and penetration behavior, with melt pool thermophysical properties acting as secondary factors that modulate melt pool geometry, the relative frequency of pore-formation mechanisms, and subsequent pore evolution. To the best of our knowledge, this study is among the first to directly compare, by in-situ X-ray imaging, the melt pool geometric characteristics of two LP-DED materials over a broad range of identical energy density conditions. The results further demonstrate that even pronounced differences in thermophysical properties do not alter the primary origin of large pore formation, which remains governed by particle inertia, while thermophysical effects mainly influence the melt pool size, shape, and the probability of each pore-formation pathway. This distinction provides a more physically grounded basis for understanding and controlling pore formation in powder-blown DED processes.

\section*{Acknowledgment}

This work was supported by the National Research Foundation of Korea (NRF) grant funded by the Korea government (MSIT) (IRIS RS-2025-02216260) and a grant of the Basic Research Program funded by the Korea Institute of Machinery and Materials (grant number: NK254A, Project Title: Development of Core Technologies for Advanced Chiplet Packaging Equipment). This research used resources of the Advanced Photon Source, a U.S. Department of Energy (DOE) Office of Science User Facility operated for the DOE Office of Science by Argonne National Laboratory (ANL) under Contract No. DE-AC02-06CH11357.  In addition, the authors would like to express their gratitude to Professors Jian Cao and Kornel Ehmann from Northwestern University for acquiring funding and providing guidance on this work. The authors also thank Dr. Shuheng Liao, Rujing Zha, Sanjana Subramaniam, and Anchen Tong at Northwestern University for their assistance in performing the experiments.



\pagebreak
\newpage
\begin{center}

\section*{Supplementary Information}
\textbf{Particle Inertia-Driven Pore Formation over Material Property Effects in Laser Powder-blown Directed Energy Deposition}
\end{center}

\setcounter{equation}{0}
\setcounter{figure}{0}
\setcounter{table}{0}
\setcounter{page}{1}
\makeatletter
\renewcommand{\theequation}{S\arabic{equation}}
\renewcommand{\thefigure}{S\arabic{figure}}
\renewcommand{\bibnumfmt}[1]{[S#1]}
\renewcommand{\citenumfont}[1]{S#1}

\subsection*{Captions for Supplementary Tables}
\noindent\textbf{Table S1} Energy density for the Ti64 and SS316 powder deposition by LP-DED\\

\noindent\textbf{Table S2} Imaging specification for the metal additive manufacturing\\

\noindent\textbf{Table S3} Input parameters used for Monte Carlo estimation of Weber number in Ti64 and SS316\\

\noindent\textbf{Table S4} Temperature-dependent thermophysical property models used for Monte Carlo estimation of Weber number\\





\newpage
\clearpage
\subsection*{Note S1. Monte Carlo estimation of the Weber number}
To evaluate the statistical variability of the Weber number under LP-DED conditions, a Monte Carlo framework was employed by propagating the uncertainty and dispersion of the governing input variables through the Weber number formulation. The analysis was intended to represent the physical variability of the LP-DED process rather than a single deterministic condition. 
In particular, particle velocity was sampled within the experimentally observed incident-velocity range obtained from high-speed X-ray measurements, whereas particle diameter was sampled from a common log-normal distribution derived from the Ti64 supplier-provided D10, D50, and D90 values and applied to both materials for comparison. In addition, melt pool temperature was treated as a bounded stochastic variable within a physically admissible LP-DED temperature range for each material.
Because the instantaneous melt pool temperature at the moment of particle impact could not be experimentally resolved, the temperature distribution was not directly assigned from measured mean and standard deviation values. Instead, temperature was sampled using a truncated normal distribution bounded by $T_{low}$ and $T_{high}$, where the mean and standard deviation were used only as distribution-shaping parameters. This treatment was intended to represent a physically realistic melt pool temperature window for Weber number estimation.

For each Monte Carlo trial, the corresponding liquid density and surface tension were evaluated from temperature-dependent property relations within the sampled temperature range. The simulation accounted for the expected variation in melt pool thermophysical properties during particle impact without assuming a single fixed melt pool temperature. For Ti64, the temperature-dependent liquid density and surface tension were evaluated using the adopted relations \cite{Mohr20, Tsen19}. For SS316, the corresponding thermophysical properties were obtained from ThermoCalc\textsuperscript{\textregistered} and subsequently linearized for use in the Monte Carlo analysis.
The detailed probability distribution functions and sampling bounds are summarized in Table S3, while the temperature-dependent thermophysical property models used in the calculation are listed in Table S4. 

All input variables were sampled independently in the present Monte Carlo analysis. A total of 100,000 samples was used for each material to ensure a stable estimation of the Weber number distribution. The lower and upper temperature bounds were selected to exclude non-physical sampling outside the melt pool temperature window relevant to LP-DED.

\newpage
\clearpage
\begin{table}[h!]
\centering
\renewcommand{\thetable}{S\arabic{table}}
\caption{Energy density for the Ti64 and SS316 powder deposition by LP-DED}
\begin{tabular}{cccccc}
\toprule
{\makecell{Material}} & {\makecell{Laser \\ power \\ (W)}} & {\makecell{Scan \\ speed \\ (mm/s)}} & {\makecell{Beam \\ dianeter \\ (mm)}} & {\makecell{Energy \\ density \\ (J/mm$^2$)}} & {\makecell{Ref.}}\\
\midrule
\multirow{16}{*}{Ti64} & 2000 & 10.6 & 4   & 47.17 & \cite{keist16} \\
                        & 300  & 4    & 1   & 75    & \cite{tan19} \\
                        & 1300 & 1300 & 2   & 0.5   & \cite{wang22(1)} \\
                        & 7600 & 15-25 & 6   & 50.67-84.44 & \cite{ren21} \\
                        & 1500-3000 & 10.67 & 4 & 35.16-70.311 & \cite{nayir21} \\
                        & 650 & 14.17 & 1.2   & 38.23 & \cite{lee23} \\
                        & 7600 & 5-25 & 6   & 50.67-253.33 & \cite{ren21(2)} \\
                        & 1000 & 11.67& 1   & 85.71 & \cite{carrozza20} \\
                        & 900-2100 & 5-15 & 3   & 20-140 & \cite{wang24} \\
                       & 30 & 10 & 0.6   & 5 & \cite{rashid22} \\
                       & 380-570 & 1000-1500 & 1.2   & 0.21-0.48 & \cite{yu12} \\
                       & 1100-1490 & 10.83-14.17 & 0.2-6   & 16.92-525.87 & \cite{qiu15} \\
                       & 2000 & 10.6 & 4   & 47.17 & \cite{carroll15} \\
                       & 600-1000 & 6.67-15 & 2   & 22.5-75 & \cite{zhang22} \\
                       & 2900 & 15 & 4.5 & 42.96 & \cite{shalnova22} \\
                       & 275 & 8.47-10.58 & 0.5   & 51.97-64.96 & \cite{yang22} \\ \hline
\multirow{15}{*}{SS316} & 1000 & 6.67-10 & 3.2 & 31.25-46.87 & \cite{benarji20} \\
                         & 450 & 3.33 & 2 & 67.51 & \cite{zhang19} \\
                         & 1000-1400 & 10 & 3.2 & 31.25-43.75 & \cite{benarji21} \\
                         & 400-800 & 5-11.67 & 0.6 & 57.14-266.67 & \cite{kumaran23} \\
                         & 600 & 3.5 & 2 & 85.71 & \cite{zhang21} \\
                         & 1300 & 16.67 & 3 & 26 & \cite{zhien21} \\
                         & 1200-2000 & 13.33 & 2 & 45-75 & \cite{feenstra20} \\
                         & 500 & 14.167 & 0.8 & 44.12 & \cite{li24} \\
                         & 1400 & 16.67 & 3 & 28 & \cite{tan19(2)} \\
                         & 750-1050 & 9.5-13 & 1.5-2 & 38.46-73.68 & \cite{sampson21} \\
                         & 600-1400 & 2-10 & 2 & 70-150 & \cite{zhang14} \\
                         & 570-750 & 12.5-16.67 & 1.2 & 37.5-38 & \cite{yu13} \\
                         & 600-1650 & 6.67-23.33 & 1 & 70.72-90 & \cite{ma13} \\
                         & 190-240 & 10-15 & 0.5 & 25.33-48 & \cite{mukherjee21} \\
                         & 750-1250 & 8.33-25 & 3 & 10-50 & \cite{lim23} \\
\bottomrule
\end{tabular}
\end{table}

\clearpage
\begin{table}[h!]
\renewcommand{\thetable}{S\arabic{table}}
\caption{Imaging specification for the metal additive manufacturing}
\renewcommand{\thetable}{S\arabic{table}}
\renewcommand{\arraystretch}{2}
\begin{tabular}{ccccccc}
\toprule
{\makecell{Manufacturing \\ technology}} & {\makecell{Powder \\ material}} & {\makecell{Frame \\ rate \\ (kHz)}} & {\makecell{Image \\ resolution}} & {\makecell{Pixel \\ size \\ ($\mu$m)}} & Multilayer & Ref.\\
\midrule
{\multirow{7}{*}{\makecell{Directed \\ energy \\ deposition}}} & {\makecell{SS316, \\ Ti-6Al-4V}} & 50 & 896$\times$448 & 2 & {\makecell{Yes \\ (2 layers)}} & {\makecell{This \\ work}} \\
                                            & {\makecell{SS316L, \\ Ti-6Al-4V}} & 80 & 1024$\times$688&  1.9        & No             & \cite{bennett22}\\
                                            & {\makecell{SS316, \\ Ti-6Al-2Sn-4Zr-6Mo}} & 2  &    -   & -  & {\makecell{Yes \\ (10 layers)}} & \cite{chen20}\\
                                            & Ti-6Al-4V         & 30 & 448$\times$388$^*$ & 2 & No               & \cite{wang22}\\
                                            & HDH Ti-6Al-4V     & 30 & 896$\times$776 & 2 & No            & \cite{wolff21}\\
                                            & \makecell{Nickel-based super \\ alloy RR1000} & 20 & 356$\times$178$^*$ & 4 & {\makecell{Yes \\ (3 layers)}} & \cite{kzhang24}\\
                                            & Cantor Alloy      & 10 & 274$\times$208$^*$& 4 & {\makecell{Yes \\ (4 layers)}} & \cite{szhang24}\\
\hline
{\multirow{5}{*}{\makecell{Laser \\ powder \\ bed \\ fusion}}}    & AlSi10Mg  & 136 & 274$\times$240$^*$  & 2 & No & \cite{hojjatzadeh19} \\
                                            & AlSi10Mg          & 50 & 384$\times$288$^*$             & 2 & No   & \cite{porter23}\\
                                            & Al7A77            & 50 & 512$\times$680 & 1.96  &  No              & \cite{huang22}\\
                                            & Ti-6Al-4V         & 20 & 1024$\times$672& 2 & No             & \cite{martin19}\\
                                            & Ti-6Al-4V         & 40 & 1008$\times$504& 4.76  & {\makecell{Yes \\ (5 layers)}}& \cite{sinclair20}\\
\bottomrule
\end{tabular}
\footnotesize{$^*$Image resolution is calculated by scale bar within a video based on the camera datasheet.}
\end{table}

\begin{table}[h!]
\renewcommand{\thetable}{S\arabic{table}}
\caption{Input parameters used for Monte Carlo estimation of Weber number in Ti64 and SS316}
\renewcommand{\thetable}{S\arabic{table}}
\footnotesize
\renewcommand{\arraystretch}{1.4}
\begin{tabular}{ccccccc}
\toprule
{\makecell{Parameter}} & {\makecell{Symbol}} & {\makecell{Ti64}} & {\makecell{SS316}} & {\makecell{Distribution}} & {\makecell{Unit}} & {\makecell{Notes}}\\
\midrule
{\makecell{Particle diameter \\ (D10)}} & $D_{10}$ & 52 & 52 & {\multirow{3}{*}{\makecell{Log-normal$^*$}}} & $\mu$m & {\multirow{3}{*}{\makecell{Ti64 datasheet; \\ same distribution \\ assumed for SS316}}} \\
{\makecell{Particle diameter \\ (D50)}} & $D_{50}$ & 71 & 71 &  & $\mu$m &  \\
{\makecell{Particle diameter \\ (D90)}} & $D_{90}$ & 102 & 102 &  & $\mu$m &  \\
\midrule
{\makecell{Velocity mean}} & $v_{\mathrm{mean}}$ & 19.3 & 15.8 & {\multirow{4}{*}{\makecell{Truncated \\ normal$^{**}$}}} & m/s & {\multirow{4}{*}{\makecell{Measured from \\ X-ray imaging}}} \\
{\makecell{Velocity standard \\ deviation}} & $v_{\mathrm{std}}$ & 4.45 & 2.59 &  & m/s &  \\
{\makecell{Velocity lower \\ bound}} & $v_{\min}$ & 0.5 & 1.2 &  & m/s &  \\
{\makecell{Velocity upper \\ bound}} & $v_{\max}$ & 29.8 & 24.2 &  & m/s &  \\
\midrule

{\makecell{Temperature lower \\ bound}} & $T_{\mathrm{low}}$ & 1923 & 1723 & {\multirow{4}{*}{\makecell{Truncated \\ normal$^{**}$}}} & K & {\multirow{2}{*}{\makecell{Physically bounded \\ LP-DED Temp. range}}} \\

{\makecell{Temperature upper \\ bound}} & $T_{\mathrm{high}}$ & 3560 & 3135 &  & K &  \\
{\makecell{Temperature mean}} & $T_{\mathrm{mean}}$ & - & - &  & K & \multirow{2}{*}{\makecell{See Appendix \\ for assumption}} \\
{\makecell{Temperature standard \\ deviation}} & $T_{\mathrm{std}}$ & - & - &  & K & \\
\midrule
{\makecell{Monte Carlo \\ sample size}} & $N$ & 100,000 & 100,000 & Fixed &  & {\makecell{Independent \\ random sampling}} \\
\bottomrule
\end{tabular}
\footnotesize{$^*$Particle diameter was modeled using a log-normal distribution based on the Ti64 powder datasheet values. $^{**}$Particle velocity and melt pool temperature were treated as truncated normal variables within physically bounded ranges.}
\end{table}

\clearpage
\begin{table}[p]
\renewcommand{\thetable}{S\arabic{table}}
\caption{Temperature-dependent thermophysical property models used for Monte Carlo estimation of Weber number}
\renewcommand{\thetable}{S\arabic{table}}
\renewcommand{\arraystretch}{1.4}
\begin{tabular*}{\textwidth}{@{\extracolsep{\fill}}cccc@{}}
\toprule
{\makecell{Material}} & {\makecell{Density, $\rho(T)$}} & {\makecell{Surface tension, $\gamma(T)$}} & {\makecell{Ref.}} \\
\midrule
Ti64 & $4300 - 0.10T$ & $2.02 - 2.5\times10^{-4}T$ & \cite{Mohr20, Tsen19} \\
SS316 & $8100 - 0.63T$ & $0.47 + 5.0\times10^{-4}T$ & {\makecell{ThermoCalc\textsuperscript{\textregistered}}} \\
\bottomrule
\end{tabular*}
\footnotesize{The temperature-dependent liquid density and surface tension were approximated using first-order linear fits extracted from the reported thermophysical property trends.}
\end{table}

\end{document}